\documentclass[aps,prl,superscriptaddress,twocolumn]{revtex4-1}
\usepackage{bbm}
\usepackage{mathrsfs}
\usepackage{amsmath}
\usepackage{amsfonts}
\usepackage[colorlinks=true,linkcolor=blue,urlcolor=blue,citecolor=blue,anchorcolor=blue]{hyperref}
\usepackage{graphicx,epstopdf}
\usepackage{subfigure}
\usepackage{epsfig}
\usepackage{dcolumn}
\usepackage{bm}
\usepackage{color}
\usepackage{natbib}
\usepackage{amssymb}
\usepackage{xcolor}
\usepackage{braket}
\usepackage{ulem}
\usepackage{float}
\usepackage{lipsum}

\definecolor{newtxtcolor}{rgb}{0.00, 0, 0}
 \newcommand { \newtxt }[1] {{\color{newtxtcolor}{#1}}}

\begin{document}

\title{ Nonreciprocal Amplification Transition in a Driven-Dissipative Quantum Network}

	\author{Mingsheng Tian}
	\thanks{M. Tian and F. Sun contributed equally to this work.}
	\address{State Key Laboratory for Mesoscopic Physics, School of Physics, Frontiers Science Center for Nano-optoelectronics, $\&$ Collaborative Innovation Center of Quantum Matter, Peking University, Beijing 100871, China}
	\author{Fengxiao Sun}
	\thanks{M. Tian and F. Sun contributed equally to this work.}
	\address{State Key Laboratory for Mesoscopic Physics, School of Physics, Frontiers Science Center for Nano-optoelectronics, $\&$ Collaborative Innovation Center of Quantum Matter, Peking University, Beijing 100871, China}
	\author{Kaiye Shi}
	\address{Department of Physics, Renmin University of China, Beijing 100872, China}
	\author{Haitan Xu}
	\email{xuht@sustech.edu.cn}
	\address{Shenzhen Institute for Quantum Science and Engineering, Southern University of Science and Technology, Shenzhen 518055, China}
    \address{School of Physical Sciences, University of Science and Technology of China, Hefei 230026, China}
	\author{Qiongyi~He}
	\email{qiongyihe@pku.edu.cn}	
	\address{State Key Laboratory for Mesoscopic Physics, School of Physics, Frontiers Science Center for Nano-optoelectronics, $\&$ Collaborative Innovation Center of Quantum Matter, Peking University, Beijing 100871, China}
	\address{Collaborative Innovation Center of Extreme Optics, Shanxi University, Taiyuan, Shanxi 030006, China}
	\author{Wei Zhang}
	\email{wzhangl@ruc.edu.cn}
	\address{Department of Physics, Renmin University of China, Beijing 100872, China}
	\address{Beijing Academy of Quantum Information Sciences, Beijing 100193, China}

\begin{abstract}
{
We study the transport properties of a driven-dissipative quantum network, where multiple bosonic cavities such as photonic microcavities are coupled through a nonreciprocal bus with unidirectional transmission. For short-range coupling between the cavities, the occurrence of nonreciprocal amplification can be linked to a topological phase transition of the underlying dynamic Hamiltonian. However, for long-range coupling, we find that the nonreciprocal amplification transition deviates drastically from the topological phase transition. Nonetheless, we show that the nonreciprocal amplification transition can be connected to the emergence of zero-energy edge states of an auxiliary Hamiltonian with chiral symmetry even in the long-range coupling limit. We also investigate the stability, the crossover from short to long-range coupling, and the bandwidth of the nonreciprocal amplification. Our work has potential application in signal transmission and amplification, and also opens a window to non-Hermitian systems with long-range coupling and nontrivial boundary effects. 
}

\end{abstract}

\maketitle
\textit{Introduction.$-$}
Nonreciprocity breaks the invariance of transmission amplitudes under the exchange of source and detector, and is of great value in numerous circumstances~\cite{review2012}. For instance, it offers new functionalities to photonic networks~\cite{zoller2017nature,graph2015,zoller2015pra,metelmann2018}, enhances the information processing capacity~\cite{opticaltransistors2010,isolator2013,zoller2019npj}, and acts as a resource for quantum metrology~\cite{clerknc2017-sensing}. A number of strategies have been developed to realize nonreciprocity, such as magneto-optical effects~\cite{mag-opt2005}, nonreciprocal phase response in Josephson circuits~\cite{PhysRevX.3.031001,PhysRevX.5.041020},  spatial-temporal refractive index modulations~\cite{RN922, PhysRevLett.109.033901}, and optomechanically induced nonreciprocity~\cite{PhysRevLett.102.213903,opome-nonre2017,RN923, RN924,xu2016,xu2019}.
Of particular interest is an amplifier with nonreciprocity which protects weak signals against noises from the read-out electronics~\cite{RevModPhys.82.1155}, and can be achieved by reservoir engineering with interfering coherent and dissipative processes~\cite{PhysRevX.5.021025, PhysRevA.95.013837,PhysRevLett.112.133904, PhysRevLett.120.023601,opome2016-np,fang2017, RN926, PhysRevX.7.031001, RN927, PhysRevApplied.11.034027}. 

Recent works show that a correspondence between nonreciprocal amplification and topological phase can be established~\cite{prl2019-topoamp, andreas-nc2020}. The transport properties of a driven-dissipative system can be characterized by a topological invariant~\cite{PhysRevLett.102.065703, PhysRevX.8.031079}. \newtxt{Since the topological characterization of a system is mostly formulated in the context of condensed matter physics where the interaction is usually short-ranged}, its application in understanding nonreciprocal amplification was focused on systems with short-range coupling~\cite{andreasprl2021, prl2019-topoamp}. However, for a driven-dissipative system possessing long-range coupling, the discussion from a topological perspective is still lacking. 

In this paper, \newtxt{we study the transport properties of a driven-dissipative system where a chain of cavities supporting bosonic modes  are coupled to a common nonreciprocal bus with unidirectional transmission. The range of the effective coupling between different cavities is determined by the coherence length of the bus, which can be much longer than the spacing between nearest neighbors.} We find that a nonreciprocal amplification can be achieved within a wide range of parameters. In the short-range coupling limit, the amplification transition can be linked to the topological phase transition either in the analysis of non-Hermitian dynamic matrix~\cite{andreas-nc2020} or within the framework of topological band theory with auxiliary chiral symmetry~\cite{prl2019-topoamp}. However, this connection fails in the long-range coupling limit, in which case the amplification transition deviates drastically from the topological phase transition associated with the dynamic matrix. This discrepancy originates from the nontrivial boundary effect induced by the long-range coupling, which breaks the conventional bulk-boundary correspondence in systems with short-range coupling. By correctly addressing the boundary effect, we analytically establish a connection between the amplification transition and the emergence of symmetry-protected zero-energy edge states for general cases with arbitrary coupling range. We further investigate the stability of the system, the crossover from short to long-range coupling, and the bandwidth of amplification.
\begin{figure}[tb]
    \centering
    \includegraphics[scale=0.15]{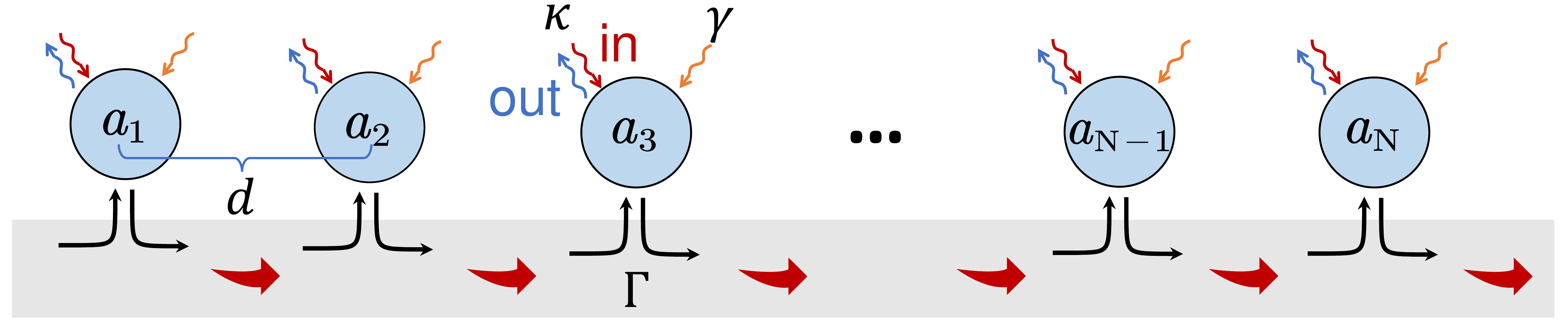}
    \caption{ \newtxt{A chain of $N$  cavities with identical bosonic modes coupled to a nonreciprocal bus. $\Gamma$ denotes the coupling between each cavity mode and the bus, $\kappa$ is the input/output coupling rate,  $\gamma$ is the net pumping rate (including internal damping), and $d$ is the spacing between neighboring cavities.}}
    \label{fig1}
\end{figure}

\textit{Nonreciprocal Amplification Transition.$-$}
We consider a one-dimensional  array of $N$ equidistantly separated cavities such as photonic microcavities with local driving and dissipation as shown in Fig.~\ref{fig1}. We focus on a single bosonic mode $a_j$ of the $j^{\rm th}$ cavity, and the bosonic modes of different cavities are assumed to have an identical frequency $\omega_0$.  \newtxt{The cavities are coupled to a nonreciprocal bus with a continuum of right-propagating bosonic modes}. The Hamiltonian reads $H_{\rm tot}=H_{S}+H_{R}+H_{SR}$, where $H_{S}=\hbar\omega_0 \sum_{j=1}^N a_j^{\dagger}a_j$ describes the $N$  cavities, and $H_{R}=\int d\omega_b\, \hbar \omega_b\,b^{\dagger}(\omega_b) b(\omega_b)$ is the free Hamiltonian of the bus modes.
The cavity-bus coupling can be modeled as
$H_{SR}=i\hbar \sum_j\int d\omega_b \sqrt{\Gamma/2\pi} b^{\dagger}(\omega_b)a_j e^{i\left(k_b x_j-\omega_b t\right)}+{\rm H.C.}$,
where $|\omega_b-\omega_0|\ll \omega_0$, $x_j$ denotes the position of the $j^{\rm th}$ cavity, and $\Gamma$ is the coupling rate between the cavity mode $a_j$ and the bus modes $b(\omega_b)$. As the bus modes  transmit unidirectionally from the left ($j=1$) to the right ($j=N$), the wavevector $k_b$ satisfies Re[$k_b]>0$, and can be expressed as $k_b= \omega_b/v + ik_b''$ where $v$ is the group velocity of the mode and $k_b''$ is related to the loss rate.

We focus on the characteristics of the input fields $\langle a_{j,{\rm in}}(t) \rangle$ and the output fields $\langle a_{j, {\rm out}}(t)\rangle$ through the driven-dissipative chain.
Under Born-Markov approximation, the equations of motion for the amplitudes $\langle a_j(t)\rangle$ are given by
\begin{eqnarray}
\label{eq3}
\langle \dot{a}_j(t) \rangle =&\frac{\gamma_j -\kappa_j -\Gamma -2i\omega_0}{2} \langle a_j(t) \rangle -\sqrt{\kappa_j} \langle a_{j,{\rm in}}(t) \rangle
\nonumber \\
	& -\sum_{j>l}^N \Gamma_{jl} e^{ik_{\omega} x_{jl}}\langle a_{l}(t) \rangle
\nonumber \\
	=& \sum_l{H_{jl}}\langle a_{l}(t)  \rangle -\sqrt{\kappa_j} \langle a_{j,{\rm in}}(t)  \rangle,
\end{eqnarray}
where \newtxt{$k_{\omega}=\omega/v$, $\omega$ is the signal frequency}, $x_{jl}=x_j-x_l$, and $\Gamma_{jl}=\Gamma e^{-x_{jl}/\zeta}$ with $\zeta= 1/k_b''$ being the coherence length of the bus mode. For simplicity, \newtxt{we assume that the input/output coupling rates and pumping rates are identical for all cavities}, i.e. $\kappa_j=\kappa$ and $\gamma_j=\gamma$, and we take the natural unit of $d =1$. \newtxt{As $\zeta$ varies from order $1$ to the order of system size $N$, the effective coupling between different cavities changes from short-range to long-range. Especially, in the infinitely long-range limit of $\zeta \gg N$, the $l^{\rm th}$ cavity is coupled to all cavities on the right with $j > l$}. 

\newtxt{The stability of the system is determined by the local parameters. A stable system requires that the real part of the eigenvalues $q_n$ of the non-Hermitian matrix $H$ to be negative}, since otherwise any fluctuation gets exponentially amplified. 
\newtxt{In our case, the eigenvalues are degenerate and
$
 Re(q_n)=(\gamma-\kappa -\Gamma)/2,
$
thus we conclude that the system is stable only when $\gamma<\kappa+\Gamma$}.

\begin{figure}[t]
    \centering
    \includegraphics[width=8.5cm]{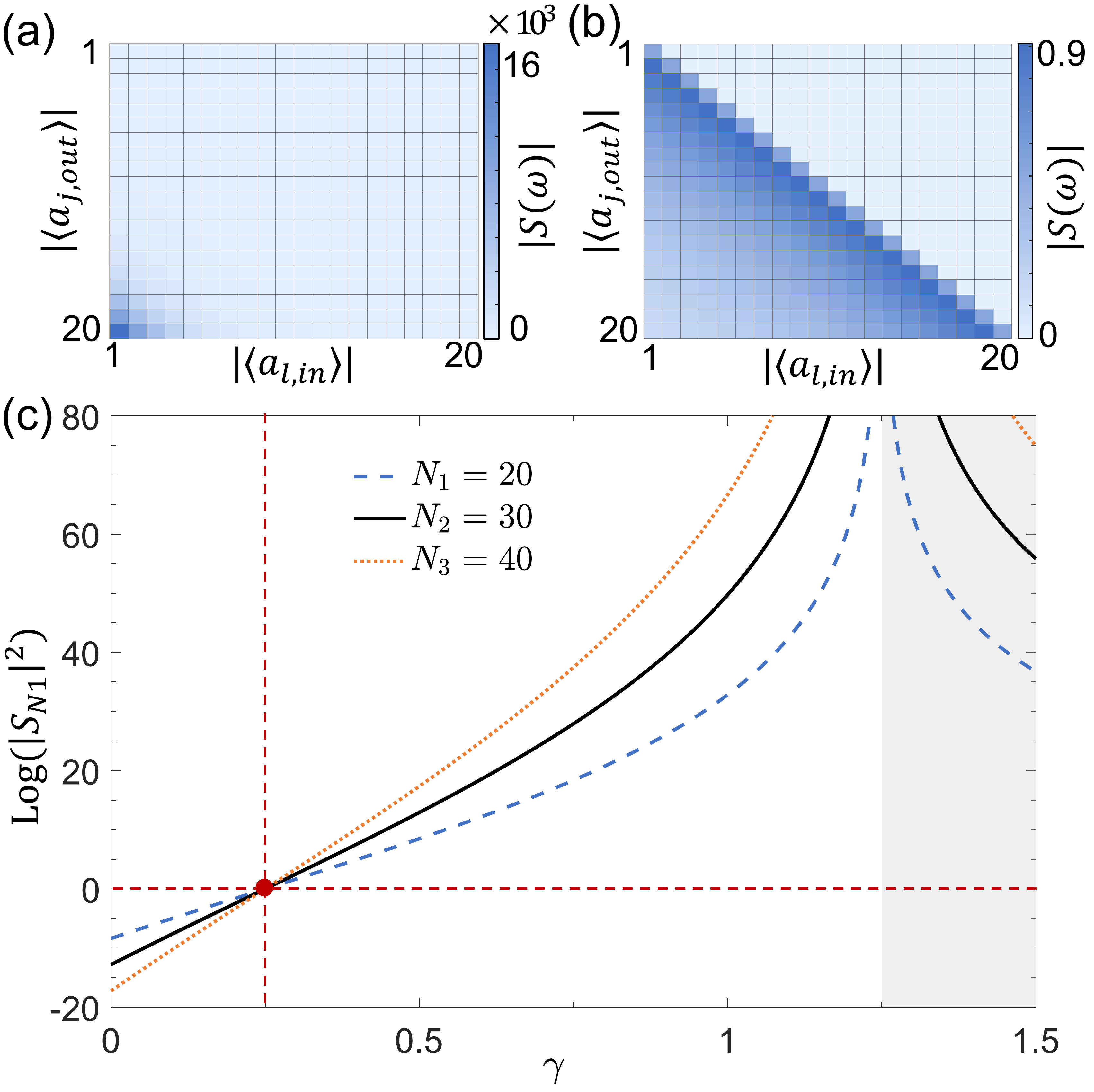}
    \caption{Scattering matrices for (a) $\gamma=0.2$ and (b) $\gamma = 0.5$ with $N=20$.
     (c) The gain $|S_{N1}|^2$ for different system sizes as a function of pumping rate $\gamma$. 
    In all plots, $\Gamma=1$, $\kappa=0.25$, $\Delta \omega=0$ and $\zeta \gg N$.
    }
    \label{fig2}
\end{figure}

By Fourier transforming $a(t)$ and considering the input-output boundary conditions $a_{j,{\rm out}}(\omega)=a_{j,{\rm in}}(\omega)+\sqrt{\kappa}a_j(\omega)$~\cite{PhysRevA.31.3761}, we obtain the input-output relation in the frequency domain
\begin{eqnarray}
\label{eq4}
\mathbf{a}_{\mathrm{out}}(\omega)
& = &\left[\mathbb{I}+\kappa(H+i (\omega-\omega_0) \mathbb{I})^{-1}\right] \mathbf{a}_{\mathrm{in}}(\omega)
\nonumber \\
&= &S(\omega) \mathbf{a}_{\mathrm{in}}(\omega),
\end{eqnarray}
where $\mathbf{a}_{\rm in/out}=(\langle a_{1,\mathrm{in/out}}\rangle,\cdots,\langle a_{N,\mathrm{in/out}}\rangle)^T$ are the input and output amplitude vectors.
The nonreciprocal amplification can be  captured by the scattering matrix $S(\omega)$. In particular, a large component in the bottom left corner of $S(\omega)$ as shown in Fig.~\ref{fig2}(a) corresponds to a strong amplification of the input signal \newtxt{from the first cavity ($j=1$) to the last one ($j=N$)}.
 
From Eq.~(\ref{eq4}), we can define a non-Hermitian dynamic matrix $M \equiv H+i(\omega-\omega_0)\mathbb{I}$ which plays an important role in characterizing the transport properties. Using the singular value decomposition (SVD) $M=U\Lambda V^{\dagger}$~\cite{PhysRevA.103.033513}, where $U$ and $V$ are unitary matrices and $\Lambda$ is a diagonal matrix $\Lambda_{nm}=\lambda_n\delta_{nm}$ with $\lambda_n>0$, we obtain an analytical solution for the scattering matrix
\begin{eqnarray}
\label{eq7}
|S_{jl}|&=&|\delta_{jl}+\kappa\sum_n V_{jn}\lambda_n^{-1} U_{ln}^*|
\nonumber \\
&=&\begin{cases}
\left|\frac{\gamma+\kappa-\Gamma+2 i \Delta\omega}{\gamma-\kappa-\Gamma+2 i \Delta\omega}\right|, & \text { for } j=l; \\
0, & \text { for } j<l;\\ 
B e^{(j-l)(\xi^{-1}-\zeta^{-1})}, & \text { for } j>l.
\end{cases}
\end{eqnarray}
The coefficients are
\begin{eqnarray}
B&=&\frac{4  \kappa \Gamma }{|\gamma+\Gamma-\kappa+2 i\Delta\omega||\gamma-\Gamma-\kappa+2 i\Delta\omega|},
\nonumber \\
\xi&=&\left [ \log \left|\frac{\gamma+\Gamma-\kappa+2 i \Delta\omega}{\gamma-\Gamma-\kappa+2 i \Delta\omega}\right| \right]^{-1},
\end{eqnarray}
where $\Delta\omega=\omega-\omega_0$ is the frequency detuning.
An amplification from the $l^{\rm th}$ cavity to the $j^{\rm th}$ cavity occurs when $|S_{jl}| > 1$. 
Especially, if the prefactor $B=1$, we have $|S_{jl}|=1$ for all $j>l$ when $\zeta=\xi$ (i.e., $\gamma=\kappa +\Gamma-\frac{2 \Gamma}{e^{1/\zeta }+1}$ \newtxt{with zero detuning}), and the amplification occurs for $\zeta>\xi$, while the attenuation occurs if $\zeta<\xi$.
For general cases with an arbitrary $B$, the same criterion of amplification holds if the separation $j-l$ is large enough. 

Of particular interest is the case of infinitely long-range coupling, i.e., $\zeta \gg N$, as it is nearly impossible in solid state systems and thus considered to be irrelevant in the context of condensed matter physics. In such case, the condition of amplification translates to $\gamma > \kappa$. This result suggests that any small cavity-bus coupling would suffice to sustain a nonreciprocal amplification provided that the pumping overcomes the dissipation to the input-output ports. As shown in Fig.~\ref{fig2}(b), the amplitudes of all scattering matrix elements are less than unity for $\gamma < \kappa$. When $\gamma$ goes beyond the transition point [red dot in Fig.~\ref{fig2}(c)], $|S_{N1}|$ becomes larger than 1 and increases rapidly with $N$, as depicted in Fig.~\ref{fig2}(c). When $\gamma$ is further increased beyond $\kappa + \Gamma$, the system enters an unstable regime labeled by shaded area in Fig.~\ref{fig2}(c). 
%

\textit{Topological invariants of dynamic matrix.$-$} The nonreciprocal amplification exhibited in the scattering matrix $S$ leaves evident signature in the non-Hermitian dynamic matrix $M$, which satisfies \textcolor{black}{$S = \mathbb{I} + \kappa M^{-1}$}. If the system has only nearest-neighbor coupling between the cavities, a direct correspondence can be established between the nonreciprocal amplification and a nontrivial topological invariant defined on the spectrum of the dynamic matrix under a periodic boundary condition (PBC)~\cite{andreas-nc2020}. However, as we will show, this correspondence does not always hold in our system. 

Under PBC, the dynamic matrix $M_{\rm{p}} = \sum_k h_{\rm{p}} (k) \vert k \rangle \langle k \vert$ is diagonal in the basis $|k\rangle=({1}/{\sqrt{N}}) \sum_j e^{ikj}|j\rangle$, where $k=2\pi n/N$ with $n=0, 1,\cdots, N-1$, and $|j \rangle$ represents the real-space wave function. The spectrum of $M_{\rm{p}}$ reads 
\begin{eqnarray}
h_{\rm{p}}(k)=\sum_{m=0}^{N-1} \mu_m e^{-ikm},
\end{eqnarray}
where $\mu_0=(\gamma-\kappa-\Gamma)/2+i\Delta\omega$ and $\mu_{m >0}=-\Gamma e^{-m/\zeta}\textcolor{black}{e^{ik_{\omega}m}}$ (see details in Supplementary Material~\cite{supplementary}).
Since $M_{\rm{p}}$ is non-Hermitian, $h_{\rm{p}}(k)=h_{{\rm p},x}(k)+ih_{{\rm p},y}(k)$ is in general complex with $h_{{\rm p}, x}, h_{{\rm p},y} \in \mathbb{R}$. Furthermore, since $h_{\rm{p}}(k)$ is periodic in $k$ with period $2\pi$, it describes a closed curve in the complex plane. Thus we can define a winding number from the argument principle~\cite{PhysRevX.8.031079,PhysRevX.9.041015},
\begin{eqnarray}
\label{eq6}
\nu_{\rm{p}} = \frac{1}{2 \pi i} \int_{0}^{2 \pi} d k \frac{h_{\rm{p}}^{\prime}(k)}{h_{\rm{p}}(k)}=
\frac{1}{2 \pi i} \int_{0}^{2 \pi} d k \frac{d}{dk} \log h_{\rm{p}}(k).
\end{eqnarray}
The winding number is an integer by counting the number of times $h_{\rm{p}}(k)$ wrapping around the origin as $k$ varies from $0$ to $2\pi$. The topological phase transition occurs at the location where the real and the imaginary parts of $h_{\rm{p}}(k)$ are both zero for some $k$. If the coupling range is much smaller than the system size ($\zeta \ll N$) and the detuning $\Delta \omega$ is zero, the transition point is determined by the condition $\kappa = \gamma +\Gamma - \frac{2\Gamma}{e^{1/\zeta}+1}$, which is consistent with the amplification transition at $\zeta = \xi$ obtained from Eq.~(\ref{eq7}). However, in the case of long-range coupling $\zeta \gg N$ and zero detuning,  the topological phase transition occurs at $\gamma = {\kappa-\Gamma}$, which is clearly different from the nonreciprocal amplification transition at $\gamma = \kappa$~\cite{supplementary}.

The breakdown of the correspondence between the topological phase transition and the amplification transition is rooted in the long-range interaction. In the limit of $\zeta \gg N$, as the bus is unidirectional, there are in total $N(N-1)/2$ coupling terms in the dynamic matrix when discussing the input-output relation under an open boundary condition (OBC). However, in the analysis of topological properties of the spectra $h_{\rm p}(k)$ under PBC, all pairs of sites are coupled and the number of coupling terms in $M_{\rm{p}}$ is doubled. Thus the difference between OBC and PBC can no longer be ignored in the thermodynamic limit. As a comparison, if the coupling is short-ranged with $\zeta \ll N$, by changing from PBC to OBC one only needs to disconnect an order of $\zeta$ couplings and induce negligible effect with large enough $N$.

\begin{figure}[t]
    \centering
    \includegraphics[width=8.5cm]{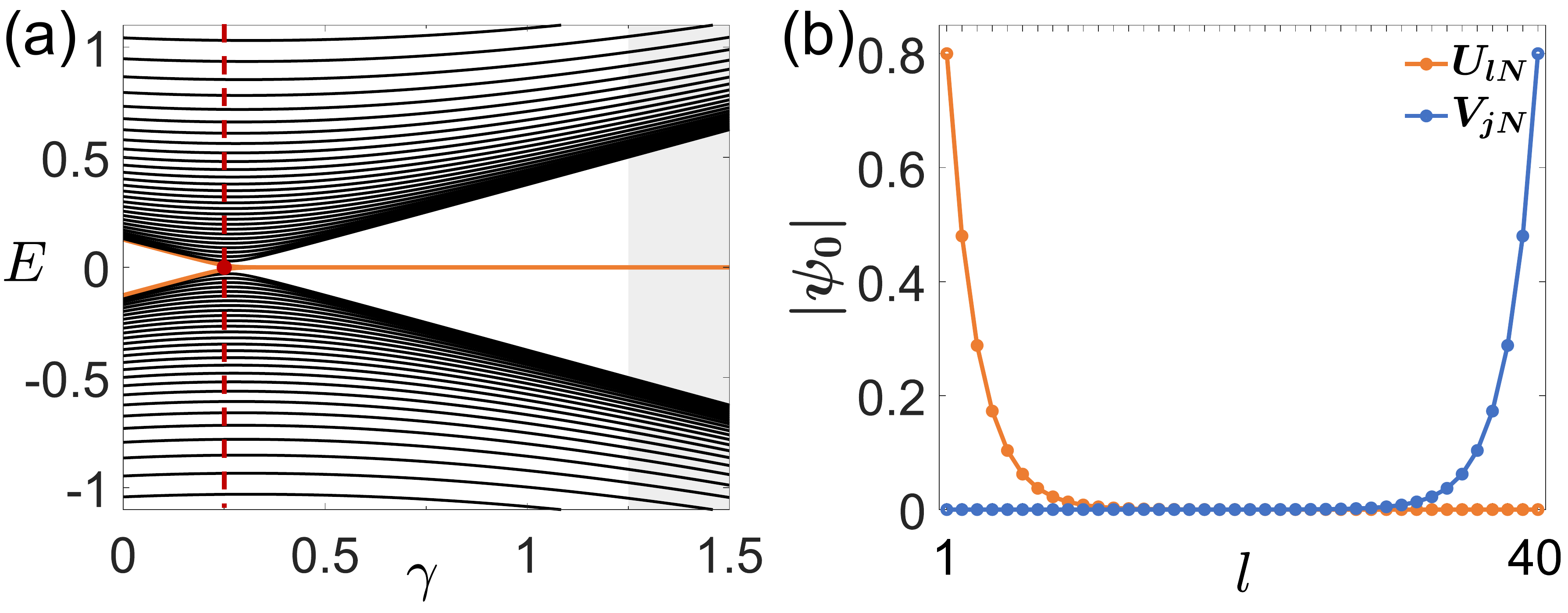}
    \caption{(a) Energy spectrum of the auxiliary Hamiltonian. A pair of zero-energy edge states (orange solid line) emerge as $\gamma$ inreases beyond a critical point (red dashed line). The unstable region is labeled by shaded area.
    (b) The amplitude of the wavefunctions $|\psi_0|$ for the zero-energy edge states with $\gamma=0.5$.
   In all plots, $N=40$, $\Gamma=1$, $\kappa=0.25$, $\Delta\omega=0$ and $\zeta\gg N$.
  }
    \label{fig3}
\end{figure}
%

\textit{From zero-energy edge states to nonreciprocal amplification.$-$}
\newtxt{Due to the drastic difference between models under OBC and PBC, we turn to the edge properties of an open boundary system to seek  further understanding of the nonreciprocal amplification transition.} 
To this aim, we define an auxiliary Hamiltonian of the system by introducing a spin and extending the dynamic matrix $M_{\rm{o}}$ 
to acknowledge the chiral symmetry~\cite{prl2019-topoamp} as
$
\mathcal{M}_{\rm{o}}=M_{\rm{o}} \otimes \sigma_{+} + M_{\rm{o}}^{\dagger} \otimes \sigma_{-},
$
where $\sigma_\pm$ are the spin ladder operators. 

\newtxt{The symmetry-protected zero-energy edge states }of $\mathcal{M}_{\rm{o}}$ emerge when 
$\zeta>\xi$ [as shown in Fig.~\ref{fig3}(a)], which coincides with the nonreciprocal amplification transition.
In fact, by directly diagonalizing $\mathcal{M}_{\rm{o}}$ in the thermodynamic limit $N \to \infty$, one can obtain an analytic result for the edge state wavefunctions 
$|U_{lN}|=N_0 e^{-(l-1)/{\xi'}}$
and 
$|V_{jN}|=N_0 e^{-(N-j)/{\xi'}}$ [see  Fig.~\ref{fig3}(b)], 
where $N_0$ is the normalization coefficient  and
$\xi'=1/(\xi^{-1}-\zeta^{-1})$
is the localization length of edge states. The corresponding eigenvalues are 
$  \lambda_{\pm}=\pm \frac{e^{-1/\zeta } (\gamma +\Gamma -\kappa )^2}{4 \Gamma } N_0^2 e^{-N/\xi'}$, which go to 0 in the thermodynamic limit
(see Supplementary Material~\cite{supplementary}).
The scattering matrix can therefore be approximated as
\begin{eqnarray}
\label{eq9}
    |S_{jl}| & \approx & \kappa|V_{jN}|\lambda_+^{-1} |U_{lN}^*|
    \nonumber \\
    &= &\frac{4 \kappa\Gamma }{|\gamma-\Gamma -\kappa|  |\gamma+\Gamma -\kappa|  }e^{(j-l)/\xi'},
\end{eqnarray}
which is consistent with that obtained from the input-output theory as given in Eq.~(\ref{eq7}) for $j>l$. We emphasize that such correspondence between the emergence of symmetry-protected zero-energy edge states and the nonreciprocal amplification transition is valid for the general case with an arbitrary coupling range.  

\begin{figure}[t]
    \centering
    \includegraphics[width=8.5cm]{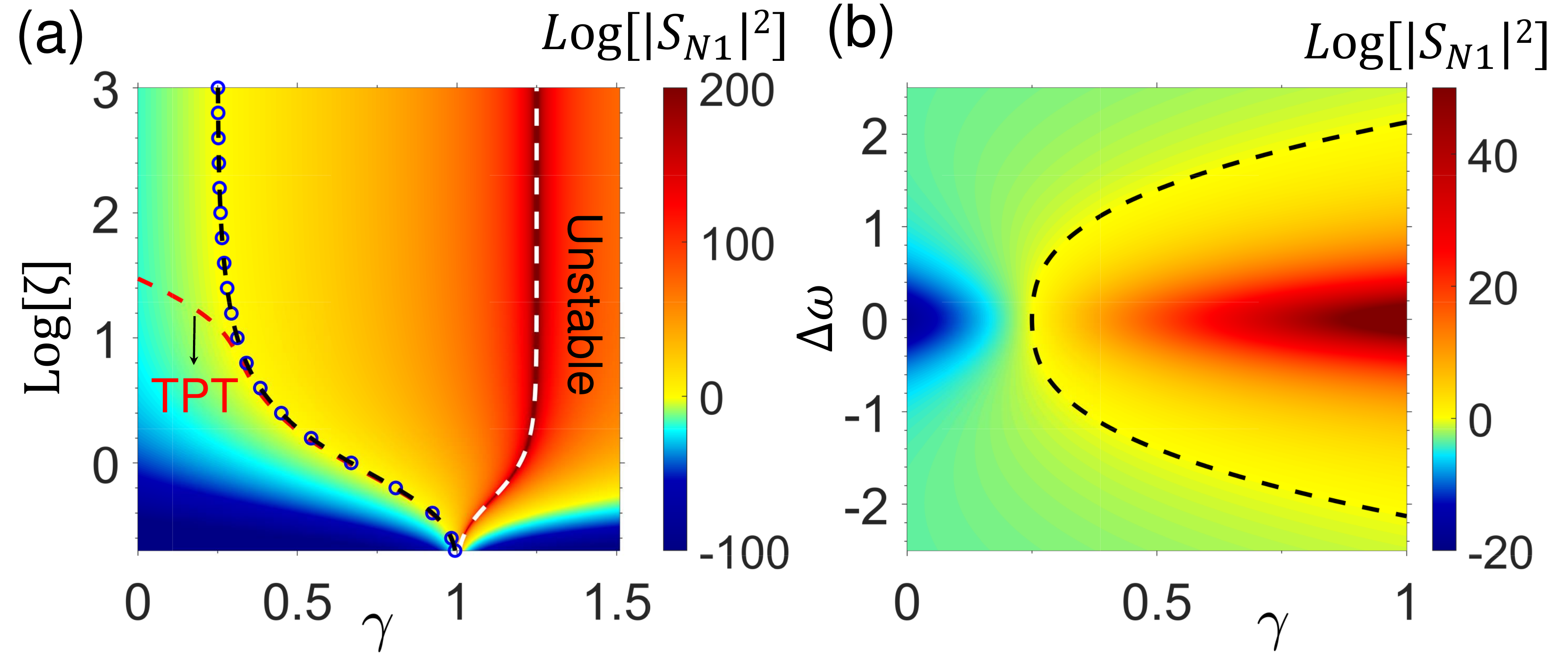}
    \caption{(a) The gain $|S_{N1}|^2$ as a function of the pumping rate $\gamma$ and the localization length $\zeta$. 
    The red dashed curve labeled by `TPT' corresponds to the topological phase transition under PBC, which agrees well with the  nonreciprocal amplification transition (black dashed curve) when $\zeta\ll N$. When $\zeta$ increases, the TPT starts to deviate from amplification transition. The emergence of zero-energy edge states (blue circles) always coincides with the amplification transition for all the range of $\zeta$. The white dashed curve indicates the critical pumping beyond which the system becomes unstable.
    (b) The gain $|S_{N1}|^2$ as a function of $\gamma$ and the frequency detuning $\Delta \omega$ for the long-range coupling limit $\zeta\gg N$. \newtxt{The black dashed curve corresponds to nonreciprocal amplification transition.} In all plots, we take $N=40$, $\Gamma=1$, and $\kappa=e^{-1/\zeta}/(1+e^{-1/\zeta})^2$. 
    }
    \label{fig4}
\end{figure}

\newtxt{In Fig.~{4(a)}, we present the gain $|S_{N1}|^2$ from the first cavity to the last by changing the pumping rate $\gamma$ and the coupling range $\zeta$. Three regions can be clearly identified, i.e.,  attenuation (left), amplification (middle), and unstable (right) regions.} The transition between attenuation and amplification (black dashed line) coincides exactly with the emergence of symmetry-protected zero-energy edge states (blue circles) for all parameters. As a comparison, the topological phase transition of the dynamic matrix $M_{\rm p}$ under PBC (red dashed line) is very close to the  amplification transition only for short-range coupling ($\zeta \ll N$), where the boundary effect is negligible~\cite{prl2019-topoamp, andreas-nc2020,PhysRevB.78.195125, Ryu_2010, asboth2016short, RevModPhys.88.035005}. With increasing $\zeta$,  the system gradually becomes a long-range coupling model and the correspondence between topological phase transition and nonreciprocal amplification transition breaks down. 
As shown in Fig.~\ref{fig4}(a), the topological phase transition starts to deviate from the amplification transition when $\zeta$ approaches a significant fraction of $N$, and becomes far off in the long-range coupling limit.


\newtxt{Furthermore, we calculate the gain $|S_{N1}|^2$ for nonzero detuning to manifest the bandwidth of amplification}. We plot in Fig.~\ref{fig4}(b) the gain $|S_{N1}|^2$ as a function of frequency detuning $\Delta \omega$ and the pumping rate $\gamma$, and conclude that the amplification bandwidth decreases with increasing $\gamma$, which is determined by Eq.~(\ref{eq7}).
Finally, we discuss possible experimental realizations of the proposed model.
Recently, nonreciprocal transport and amplification have been
observed by optomechanical interactions~\cite{opome2016-np,fang2017}.
The whispering gallery modes in microresonators can be chirally coupled with forward propagating wave in tapered fibre, which is also a feasible platform to simulate the model described by Eq.~(\ref{eq3}).
Besides, the nonreciprocal amplification can also be realized in superconducting systems~\cite{sup2014,sup-sci2015}.

\textit{Conclusion.$-$}
\newtxt{We have studied the amplification transition in a quantum network of cavities coupled to a nonreciprocal bus with unidirectional transmission. When the coherence length of the bus is comparable to or even exceeds the system size, an effective long-range coupling between the cavities can be established}. We have derived the input-output relation for this system, and obtained an analytical condition for the nonreciprocal amplification transition, which, however, cannot be directly linked to the topological phase transition associated with the spectrum of the dynamic matrix under a periodic boundary condition. In the long-range coupling limit, the topological phase transition drastically deviates from the amplification transition, owing to the significant boundary effect. To fill this gap, we  work on the system with open boundary condition and establish an exact connection between the nonreciprocal amplification transition and \newtxt{the emergence of symmetry-protected zero-energy edge states} for arbitrary coupling range. Our work can be applied to signal transmission and amplification, and also paves the way to study the long-range coupling and nontrivial boundary effects in non-Hermitian systems.


\begin{acknowledgments}
This work is supported by the National Natural Science Foundation of China (Grants No.~11974031, 11975026,  12125402,  12147148, and  12074428), Beijing Natural Science Foundation (Grant No. Z190005 and No. Z180013), the Key R$\&$D Program of Guangdong Province (Grant No. 2018B030329001 and No. 2018YFA0306501). F. S. acknowledges the China Postdoctoral Science Foundation (Grant No. 2020M680186).
\end{acknowledgments}




%


\end{document}